# An "invisible" nonvolatile solid-state memory

**Authors:** J. Clarkson[1], C. Frontera[2], Z. Q. Liu[1,3], Y. Lee[1], J. Kim[4], K. Cordero[5], S. Wizotsky[6], F. Sanchez[2], J. Sort[7,8], S. L. Hsu[1,9], C Ko[1], J. Wu[1,9], H.M. Christen[3], J. T. Heron[8,9], D.G. Schlom[8], S. Salahuddin[10], L. Aballe,[11] M. Foerster,[11] N. Kioussis[4], J. Fontcuberta[2], I. Fina[2], R. Ramesh[1,13,14] and X. Marti[15]

**Affiliations:**
[1] *Department of Materials Science and Engineering, University of California, Berkeley, California 94720, USA*

[2] *Institut de Ciència de Materials de Barcelona (ICMAB-CSIC), Campus UAB, Bellaterra 08193, Barcelona, Spain*

[3] *Oak Ridge National Laboratory, Center for Nanophase Materials Sciences, Oak Ridge, Tennessee 37831, USA*

[4] *Department of Physics, California State University, Northridge, California 91330-8268, USA*

[5] *Catalan Institute of Nanoscience and Nanotechnology (ICN2), CSIC and The Barcelona Institute of Science and Technology, Campus UAB, Bellaterra, 08193 Barcelona, Spain*

[6] *Max Planck Institute of Microstructure Physics, Weinberg 2, D-06120 Halle (Saale), Germany*

[7] *Departament de Física, Universitat Autònoma de Barcelona, E-08193 Bellaterra, Spain*

[8] *Institució Catalana de Recerca i Estudis Avançats (ICREA), E-08010 Barcelona, Spain*

[9] *Materials Sciences Division, Lawrence Berkeley National Laboratory, Berkeley, California 94720, USA*

[10] *Department of Materials Science and Engineering, Cornell University, Ithaca, New York 14850*

[11] *ALBA Synchrotron Light Facility, Carretera BP 1413, Km. 3.3, Cerdanyola del Vallès, Barcelona 08290, Spain.*

[12] *Department of Materials Science and Engineering, University of Michigan, Ann Arbor, Michigan 48109, USA*

[13] *Department of Electrical Engineering and Computer Science, University of California, Berkeley, California 94720, USA*

[14] *Department of Physics, University of California, Berkeley, California 94720, USA*

[15] *Institute of Physics ASCR, v.v.i., Cukrovarnicka 10, 162 53 Praha 6, Czech Republic*

**An electric field cloakable non-volatile solid state memory is demonstrated, promoting new secure memory and logic architectures**

**Abstract**: **Information technologies require entangling data stability with encryption for a next generation of secure data storage. Current magnetic memories, ranging from low-density stripes up to high-density hard drives, can ultimately be detected using routinely available probes or manipulated by external magnetic perturbations. Antiferromagnetic resistors feature unrivalled robustness but the stable resistive states reported scarcely differ by more than a fraction of a percent at room temperature. Here we show that the metamagnetic (ferromagnetic to antiferromagnetic) transition in intermetallic $Fe_{0.50}Rh_{0.50}$ can be electrically controlled in a magnetoelectric heterostructure to reveal or cloak a given ferromagnetic state. From an aligned ferromagnetic phase, magnetic states are frozen into the antiferromagnetic phase by the application of an electric field, thus eliminating the stray field and likewise making it insensitive to external magnetic field. Application of a reverse electric field reverts the antiferromagnetic state to the original ferromagnetic state. Our work demonstrates the building blocks of a feasible, extremely stable, non-volatile, electrically addressable, low-energy dissipation, magnetoelectric multiferroic memory.**

**Main Text:**

Contemporaneous to the discovery of the transistor, magnetic media became the prime support for the majority of long-term information storage (*1*). The need for protecting the recorded information from either sabotage or unauthorized access is now a critical requirement in current information technologies (*2*). The first step towards breaching any security system is sensing the mere presence of the information that is stored, for example, as magnetic bits. In this regard, it is very appealing to render the magnetic bits "invisible", that is, to cloak the magnetic signature of the repository. Cloaking has been proposed through modifications of optical properties in a number of meta-materials (*3-5*) and superconducting systems(*6*) but has yet to be demonstrated in a data storage technology. Recently, antiferromagnetic memory elements, that intrinsically produce null stray fields, have been shown to be capable of storing magnetic information at room temperature (*7-9*), but these "invisible" bits are measured through small resistance variations or the anomalous Hall effect in non-collinear antiferromagnets (*8, 10*). In this Report, we demonstrate a cloakable magnetic bit encoded in a single compound, operating at or above room temperature. We present an all-electrical methodology to render the magnetic information invisible and retrieve it upon demand. In addition to this novel concept, all previously available developments in data security can be built on top of our proof-of-concept.

Our device assigns the data storage and the caching action to two strongly coupled materials. First, we employ a thin layer of $Fe_{0.50}Rh_{0.50}$ (FeRh) which is an intermetallic compound that exhibits an antiferromagnetic (AFM) to ferromagnetic (FM) phase transition that can be controlled by strain, heat, or magnetic field (*11-16*). Concurrent with its metamagnetic phase transition, FeRh displays a very large change in resistivity (~30-40%) (*11*) which can be triggered by any of the aforementioned stimuli. Second, our device is comprised of a piezoelectric substrate (for example, $(0.72)PbMg_{1/3}Nb_{2/3}O_3$ - $(0.28)PbTiO_3$ (PMN-PT) or $BaTiO_3$) which allow the tuning of the strain exerted on our magnetic material. This combination of materials has achieved, in preceding works, an electrically driven modulation of magnetization(*15*, *17*) by as large as 70 emu/cm$^3$ (~21% isothermal magnetization modulation) (*15*) and resistivity modulation of 8-22% (*16, 18*) thus proving the magnetoelectric coupling of the heterostructure. Specific details for the very same samples employed in this work are found elsewhere (*16*, *18*).

Our model system consists of a 30 nm thick FeRh layer that is deposited by ultrahigh vacuum sputtering onto a single crystal of PMN-PT at a deposition temperature of 350 °C. To demonstrate the generality of this approach, a single crystal $BaTiO_3$ was also used as a piezoelectric substrate (details presented in the supplementary section). Prior to our magnetoelectric studies, we characterized the crystalline quality and magnetic properties of the FeRh layer (Supplementary S1-2). Armed with this baseline, we now present results of temperature and electric field dependent magnetic and magnetotransport studies that form the fundamental basis for the "cloaking" protocol. Figure 1A presents temperature dependent SQUID magnetic hysteresis loops for the FeRh, measured in-plane and out-of-plane at 27 °C and 127 °C. Upon lowering the temperature two changes occur. First, the absolute magnetic moment is reduced, consistent with the emergence of the antiferromagnetic phase. Second, and perhaps more importantly, the magnetization is now directed significantly more out-of-plane, as illustrated in the comparison of the in-plane versus out-of-plane magnetic hysteresis loops. This out-of-plane component of the magnetization is key to detectability of this state (in our case through anomalous Hall resistance (*19*, *20*) [AHR, $R_{xy}$], described below).

4-probe transport measurements were carried out as a function of temperature and electric field (S3). In the mixed AFM-FM state (for example at a temperature of 60 °C) application of an out-of-plane electric field to the heterostructure leads to a strong change in the (longitudinal) resistance, Figure 1B, which is directly attributed to the electric field control of the magnetic state (FM vs. AFM) (*15*, *16*, *18*). Positive (negative) electric fields produce an increase (decrease) of electrical resistance, corresponding to an increase in the AFM (FM) phase fraction due to the strain-induced preference for this phase(*15*, *16*, *18*). Using this as the starting point, we now describe how such a heterostructure can be used to cloak the written magnetic state. The strategy we propose is schematically summarized in Figure 1C. The memory element is programmed into the UP and DOWN states, [step 1 and 1'] with a magnetic field; a subsequent electric field [step 2 and 2'] converts it into the AFM phase thus cloaking and protecting it from external perturbation. An electric field of opposite polarity [step 3 and 3'] converts it back to the original FM state. The main experimental result to validate this protocol is summarized in Figure 1D which shows the AHR of the FeRh layer as a function of the electric field applied to the underlying piezoelectric. After magnetizing the FeRh into the UP (or DOWN) states, a negative electric field, favors ferromagnetism (state 1 and 1'), while a positive field converts it into the AFM state with a lower AHR (states 2 and 2'), which are essentially the same.

In order to understand the effects of in-plane strain and of the relative volume fraction of the FM phase on the magnetic moment orientation, we have carried out *ab-initio* electronic structure calculations of the volume-averaged magnetic anisotropy energy (MAE), $\overline{K}_V(\varepsilon, x) = (1-x)K_V^{AFM}(\varepsilon) + x\,K_V^{FM}(\varepsilon)$. Here, $\varepsilon = (a_o^{FM} - a_o^{AFM})/a_o^{FM}$,is the strain relative to the equilibrium FM-FeRh lattice constant, $a_o^{FM}$, $K_V^{AFM(FM)}(\varepsilon)$is the strain dependent intrinsic bulk MAE for the AFM (FM) FeRh phase and $x = \frac{V^{FM}(\varepsilon)}{V^{FM}(\varepsilon)+V^{AFM}(\varepsilon)}$ is the relative volume fraction of the FM phase. In order to investigate the effect of magnetic shape anisotropy energy (MSAE) on the FM thin film, we have also calculated the effective MAE, $K_{eff}^{FM}(\varepsilon) = K^{FM}(\varepsilon) - 2\pi M^2(\varepsilon, x)$, where $M\,(\varepsilon, x) = x(m_{Fe}(\varepsilon) + m_{Rh}(\varepsilon))/V(\varepsilon)$, $m_{Fe}(\varepsilon)$, $m_{Rh}(\varepsilon)$, and $V(\varepsilon)$, are the weakly strain-dependent magnetic moments of the Fe and Rh atoms, and the volume of the formula unit cell, respectively.

In Figure 2A we show the variation of the volume-averaged MAE, $\overline{K}_V$ with strain and relative volume fraction of the FM phase. We find that the MAE is positive (negative) under compressive (tensile) strain for the FM phase, indicating an out-of-(in-)plane orientation of the magnetic moments. In sharp contrast, the strain dependence of the MAE in the AFM phase is reversed, consistent with previous results (*21*). The variation of the effective MAE, $K_{eff}^{FM}$, of the FM phase is shown in Figure 2B. Interestingly, for x = 1, the MSAE renders $K_{eff}^{FM} < 0$ in the entire strain range, demonstrating that when the entire volume is in the FM phase, the spin orientation is in-plane. As the relative volume fraction of the FM phase decreases, the effect of MSAE also decreases and $K_{eff}^{FM}$ is determined primarily by $K_V^{FM}$ yielding an out-of-plane spin orientation under compressive strain.

Supporting evidence for the electric field modulation of the magnetic state described in Figure 1D comes from analogous temperature dependent programming of the magnetic state, as shown in Figure 3A, which shows a sequence of thermal programming steps in which the FeRh layer is oriented in a magnetic field of +(-) 500 mT, then cooled from the FM state (state 1 and 1') at 127

°C in to an essentially AFM state at 27 °C (state 2 and 2'). Upon heating back to 127 °C, the same ferromagnetic polarity is retrieved (i.e. state 3 and 3'), suggesting the existence of FM kernels within the AFM phase, which contain the directional information and assist in attaining the originally written FM state upon heating. Figure 3B demonstrates that such a reversal can be carried out over several cycles. The corresponding electric field cycling results are presented in Figure 3C which shows the AHR as a function of time at positive, zero and negative magnetic fields as a function of electric field cycling at 60 °C. It is clear that the distinguishability of the two magnetic states is electric field polarity dependent and stable over several cycles. The horizontal line in Figure 3C indicates the AFM or "cloaked" state, while the data significantly above or below the line represent a positive or negative magnetization state. Through a negative applied electric field the magnetization direction of this "cloaked" state is revealed. To establish the robustness of this data against external magnetic fields we deliberately probed the magnetic field dependence of the AHR in both the FM and AFM states at zero electric field, as shown in Fig. 3D. The fact that the magnetic field dependence for the two states do not cross or join implies that the AFM state cannot be "broken" into by an external magnetic field, at least up to 300 mT at a temperature of 27 °C.

We now probe the microscopic details of such a directional magnetic memory, using magnetic force microscopy (MFM) as a function of applied electric field. We observed measurable changes in the image contrast upon the application of positive and negative electric fields. Figure 4A, obtained with a positive electric field of 7.3 kV/cm shows an essentially "uniform" green contrast, indicative of an essentially antiferromagnetic state (i.e., no significant magnetic moment); upon the application of a -7.3 kV/cm electric field (Figure 4B), the image contrast changes with the emergence of regions with strong contrast (in blue and red), indicative of the emergence of the ferromagnetic state. It is particularly important to note that the contrast for a given location in the sample is reversibly changed from red (or blue) to green, again providing supporting evidence to the notion that a specific ferromagnetic polarity is retained in that location. It has been earlier reported that the FM state set at high temperature can be imprinted in the AFM state at room temperature by fixing the direction of the antiferromagnetic axis by in-field magnetic cooling from the high temperature FM phase down to the low temperature AFM phase (*7*). Therefore, the observed memory effect can be attributed to a reciprocal effect, in which the magnetic orientation induced in the AFM state drives the magnetization orientation in the FM state upon subsequent warming (or the application of a negative electric field). The role of localized structural defects in the film structure is likely to be a key element that is required for such a memory effect. Indeed, highly epitaxial films on single-crystalline MgO substrates do not show a sizeable retained magnetic moment at room temperature and, in agreement with prior work (*22*), we observe no memory effect. Therefore, we conclude that the coexistence of FM/AFM phases at room-temperature is required to store the magnetic information and that the minority FM domains present at room temperature must play a crucial role in the observed memory effect. The coexistence of magnetic phases has been long exploited in the well-known magnetic shape-memory alloys that show magnetic memory while crossing the austenite to martensite phase transition at which the magnetic state changes (*22*, *23*). It appears that a similar phenomenology is at play here, with the advantage that the room-temperature AFM phase can mask the coexisting FM nuclei. It is worth noting that, in contrast to most common situations where single crystalline substrates are a demand for optimal thin film properties, here poly-domain substrates are definitely advantageous and thus we expect to observe similar memory effects in a wide range of inexpensive substrates. Finally, exploiting memory effects such as those described here for FeRh require materials that display near room temperature AFM to FM transitions. Fortunately several alloys, such as $Mn_2Sb$ and related compounds (*24-26*) are already known. Thus the applicability of the cloaking memory concept in a single-phase material electrically or thermally controlled at room temperature is ensured.

**Acknowledgments:**

The work at Berkeley was supported in part by FAME, one of six centers of STARnet, a Semiconductor Research Corporation program sponsored by MARCO and DARPA was supported by the SRC-FAME program. The work at ORNL was sponsored by the Laboratory Directed Research and Development (LDRD) Programs of ORNL managed by UT-Battelle, LLC. Financial support by the Spanish Government [Project MAT2014-56063-C2-1-R] and Generalitat de Catalunya (2014 SGR 734) is acknowledged. ICMAB-CSIC authors acknowledge financial support from the Spanish Ministry of Economy and Competitiveness, through the "Severo Ochoa" Programme for Centres of Excellence in R&D (SEV- 2015-0496). IF acknowledges Juan de la Cierva – Incorporación postdoctoral fellowship (IJCI-2014-19102) from the Spanish Ministry of Economy and Competitiveness of Spanish Government. XMCD-PEEM experiments were performed at CIRCE beamline at ALBA Synchrotron with the collaboration of ALBA staff. Partial financial support from the 2014-Consolidator Grant "SPIN-PORICS" (Grant Agreement nº 648454) from the European Research Council is acknowledged.

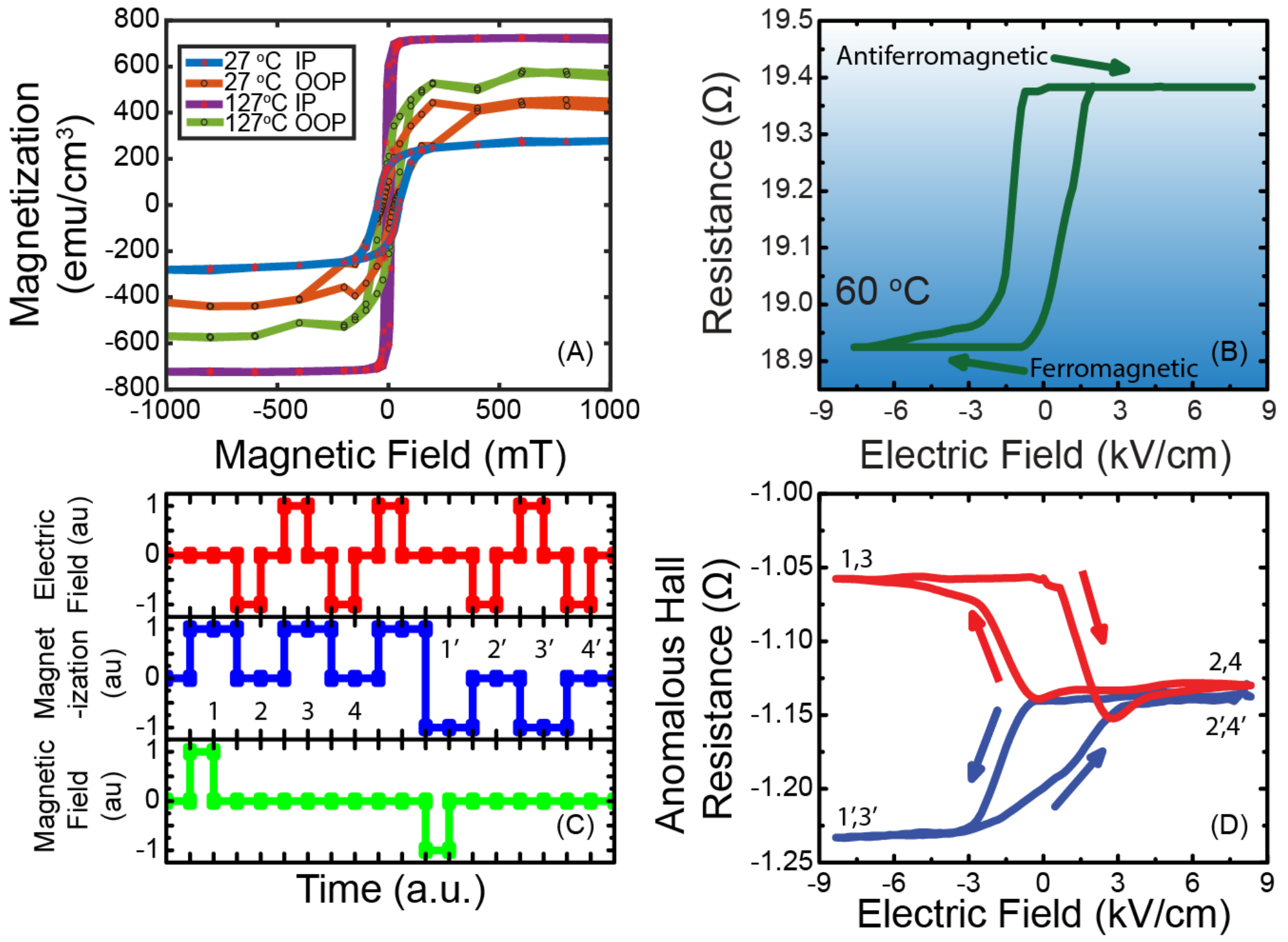

**Fig. 1:** (A) In-plane and out-of-plane magnetization hysteresis loops measured by Superconducting Quantum Interference Device (SQUID) at 27 °C and 127 °C. (B) Electric field dependence of the longitudinal resistance of FeRh (30 nm)/PMN-PT. (C) A step-wise illustration of how the memory element is programmed into a magnetic state (step 1 and 1'), "cloaked" by the application of an electric field (step 2 and 2') and "uncloaked" by the application of an opposite polarity electric field (step 3 and 3') of our proposed cloaked magnetic memory device functions. (D) Electric-field dependences of the anomalous Hall resistance (AHR) for the two magnetic states.

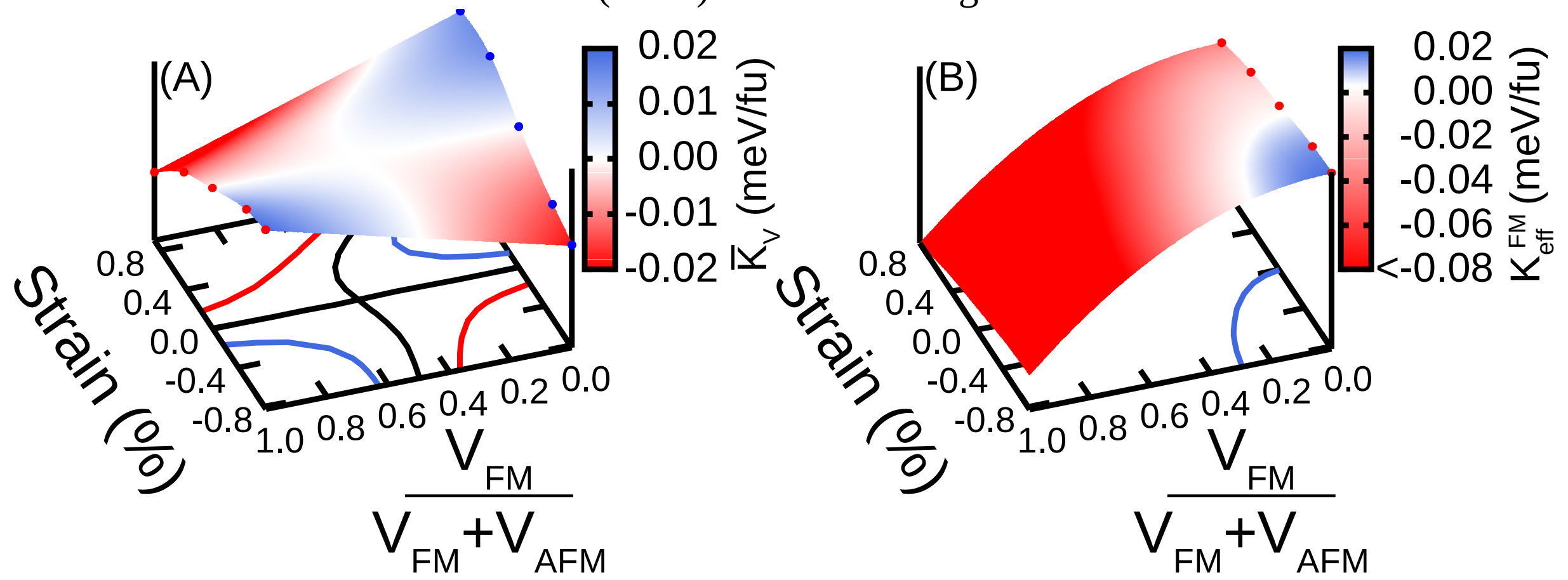

**Fig.** 2: *Ab initio* calculations results of the FeRh magnetic anisotropy as a function of strain. (a) The volume-averaged magnetic anisotropy energy, $\overline{K}_V$, as a function of strain and FM/AFM volume fraction. (b) The net magnetic anisotropy energy of the FM phase is shown, including demagnetization effects.

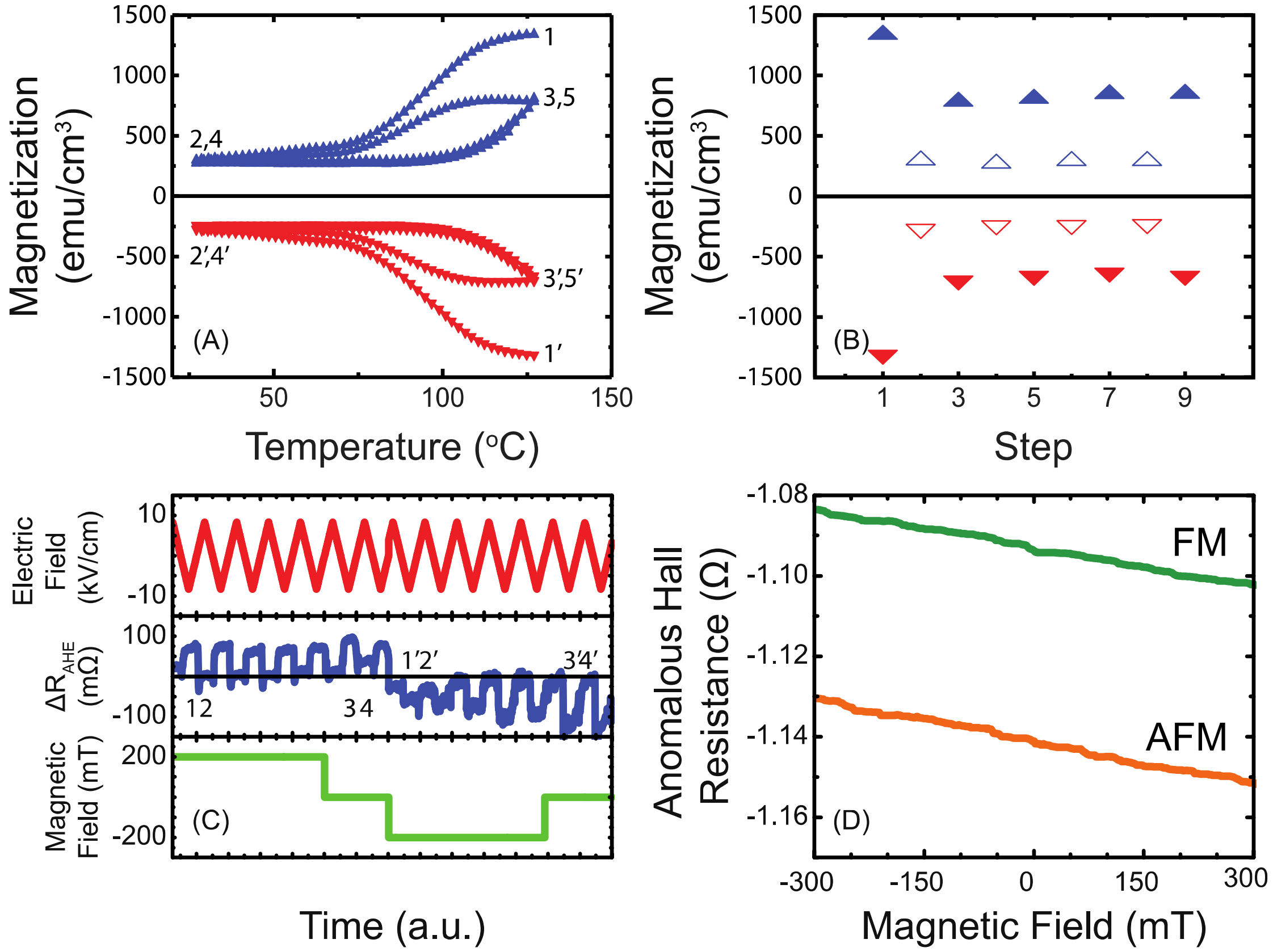

**Fig. 3:** (A) Magnetization versus temperature loops measured at 0 mT, after saturating the sample (± 500 mT) at 127 °C. Arrows indicate increasing or decreasing temperature. (B) Magnetic moment at 127 °C (solid symbols) and 27 °C (empty symbols) measured after successive thermal cycling at 0 mT. (C) Anomalous Hall resistance as a function of time (at 60 °C) at positive (200 mT), zero, and negative (-200 mT) magnetic fields through a series of different voltage cycles. Between the 0 and -200 mT fields, a field cooling was carried out. (D) AHR as a function of magnetic field for the two 0 electric-field states at 25 °C. The magnetic response is due to the ordinary Hall effect, and are separated due to differing magnetization states, AFM and FM.

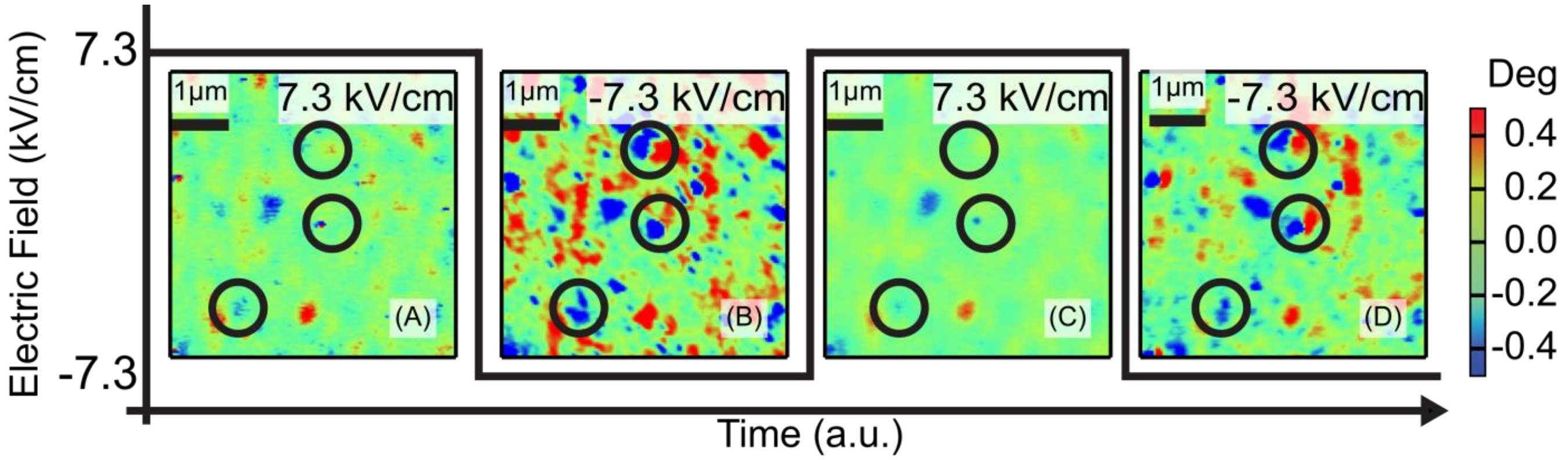

**Fig. 4:** Magnetic force microscopy of the FeRh layer at 60 °C under a DC electric field of ±7.3 kV/cm is applied. Images are labeled sequentially with their corresponding electric field.

The magnetic response in (A) and (C) are drastically diminished, when compared with the same region in (B) and (D). The local magnetic memory effect is observed here as the local magnetization is reversibly modified with electric field.

**Supplementary Materials:**

**1. X-ray diffraction and TEM of FeRh deposited on PMN-PT and $BaTiO_3$ substrates.**

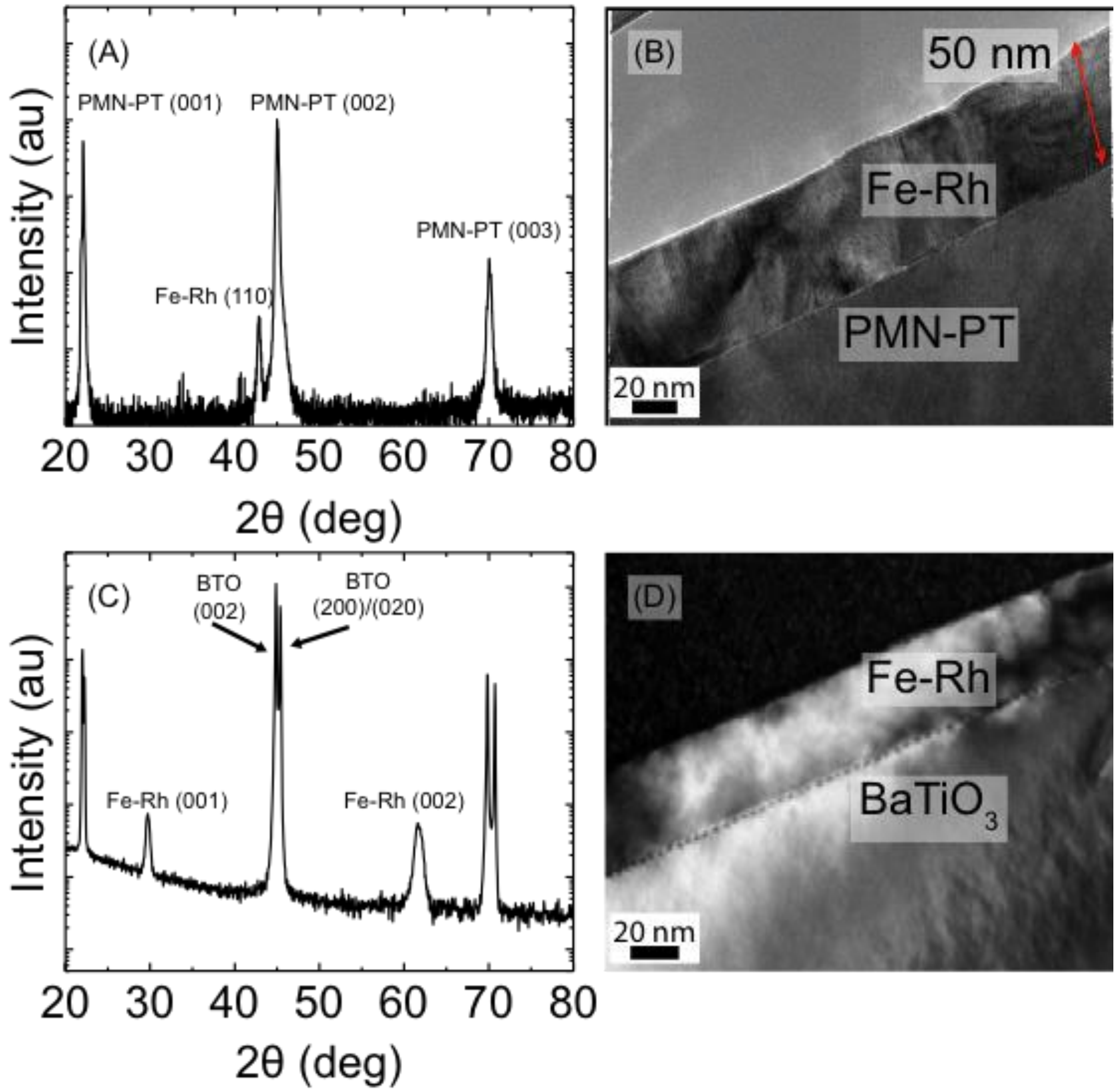

**Fig. S1:** (A) Room temperature out-of-plane x-ray diffraction, 2θ/θ, of FeRh/PMN-PT exhibiting a preferred (110) out-of-plane orientation. (B) TEM of a 50 nm FeRh film on PMN-PT. The film exhibits multiple grains. (C) Room temperature out-of-plane x-ray diffraction, 2θ/θ, of FeRh/$BaTiO_3$ exhibiting a (001) out-of-plane orientation. (D) TEM of an 50 nm FeRh film on $BaTiO_3$. No grain boundaries are observed, confirming a single crystal thin film with high crystallinity. Scale bars in (B) and (D) are 20 nm.

## 2. Magnetization as a function of temperature

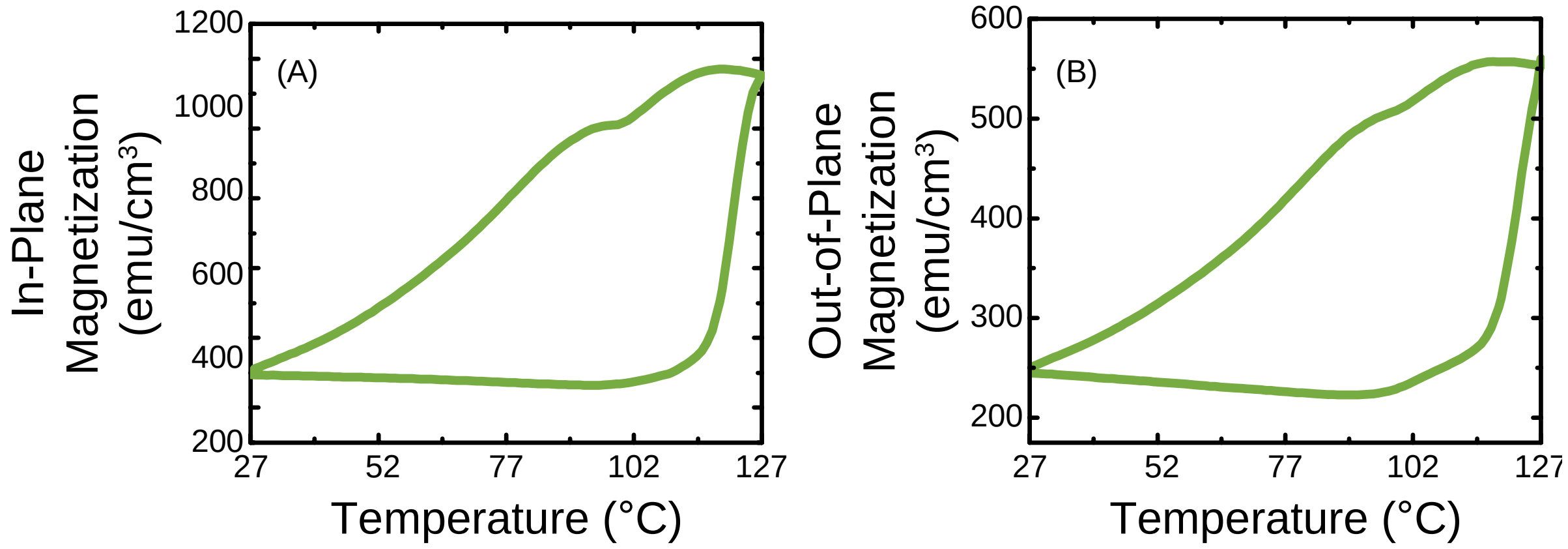

**Fig. S2:** (A) In-plane and (B) out-of-plane magnetization of a 50 nm FeRh film on PMN-PT as a function of temperature at 500 mT.

**3. Geometry of transport measurements and voltage application**

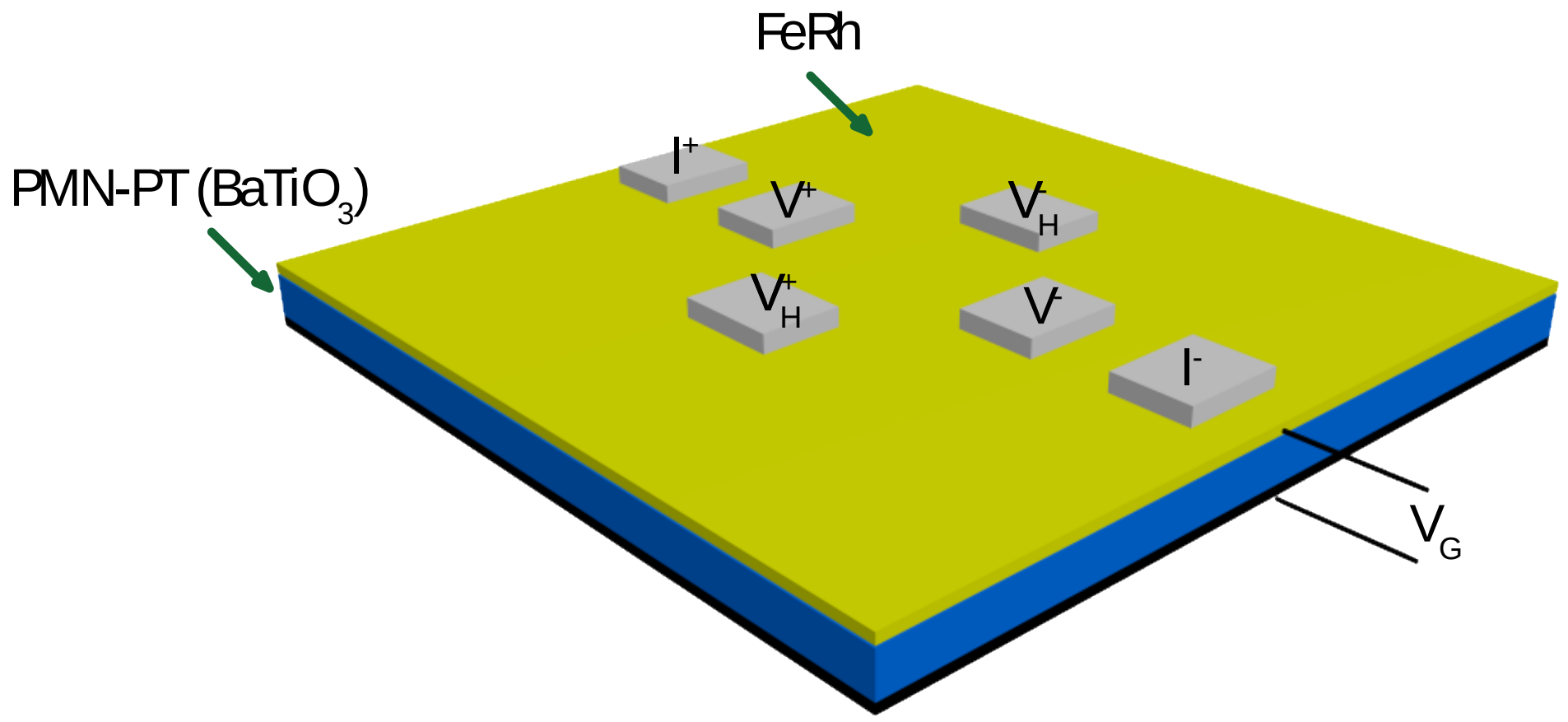

**Fig. S3:** Schematic showing the four-wire longitudinal and transverse resistivity measurement geometries for FeRh thin films. The applied magnetic field, when applied, is oriented perpendicular to the plane of the film. The voltage applied to PMN-PT substrate is done using the blanket FeRh film and a back electrode.

**4. Magnetic Anisotropy Measurements of FeRh/PMN-PT**

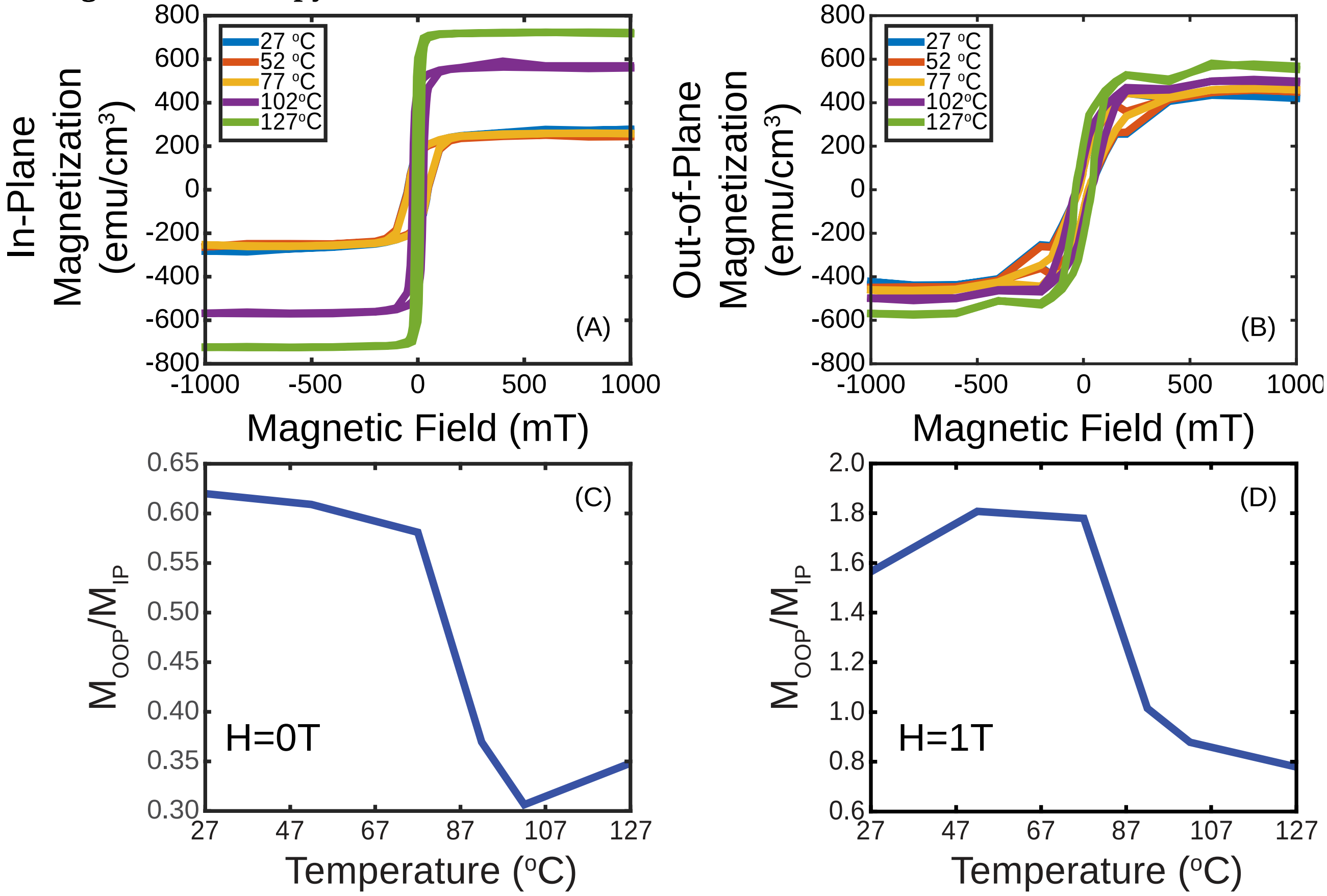

**Fig. S4:** (A) In-plane and (B) out-of-plane magnetization hysteresis loops at various temperatures across the magnetic phase transition for FeRh/PMN-PT. Ratios of magnetization components at (C) zero magnetic field and (D) 1 T, as a function of temperature, depicting the magnetic anisotropy as a function of temperature for FeRh/PMN-PT. An out-of-plane magnetization direction is preferred below the transition temperature.

**5. Magnetic Anisotropy Measurements of FeRh/PMN-PT**

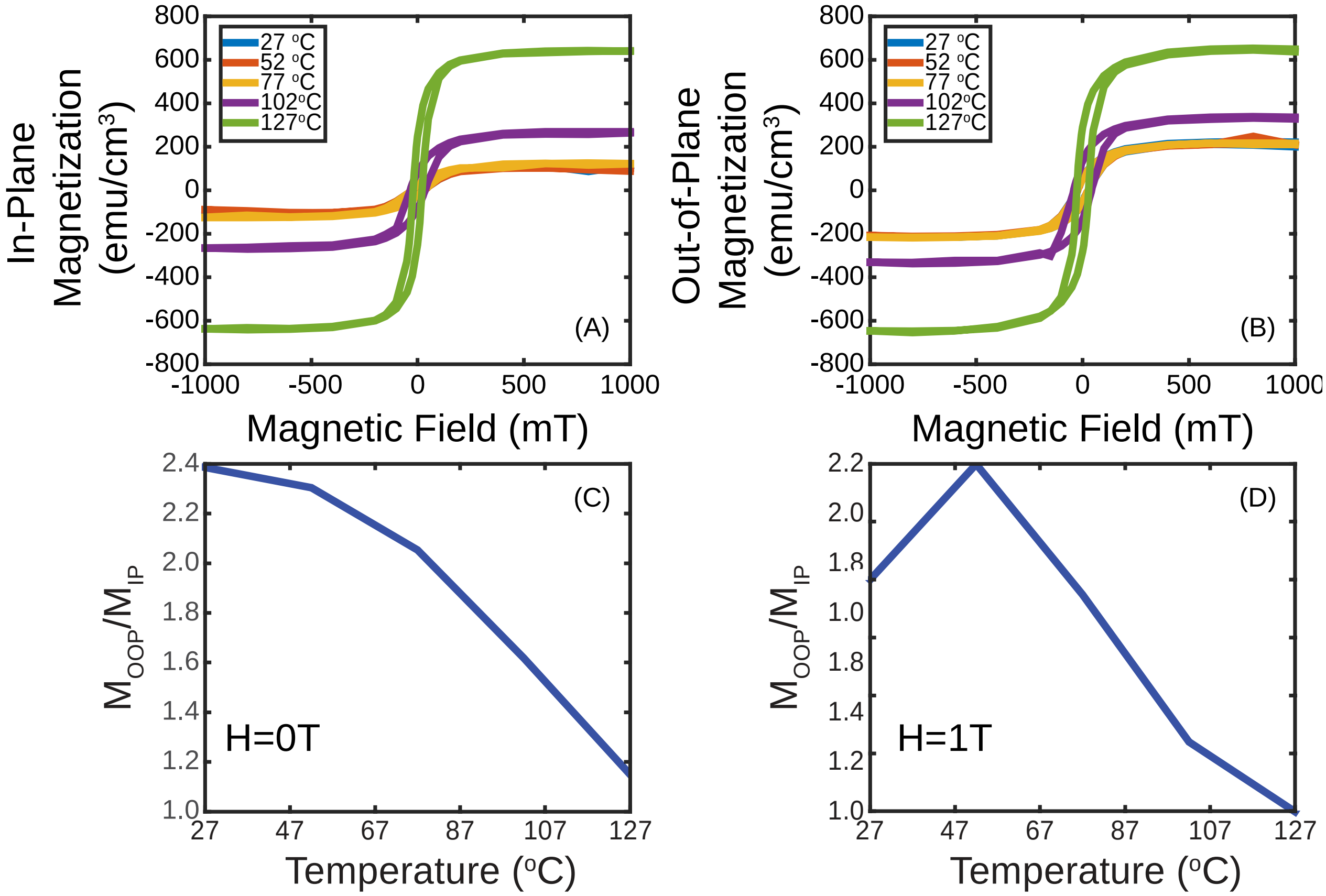

**Fig. S5:** (A) In-plane and (B) out-of-plane magnetization hysteresis loops at various temperatures across the magnetic phase transition for FeRh/ $BaTiO_3$. Ratios of magnetization components at (C) zero magnetic field and (D) 1 T, as a function of temperature, depicting the magnetic anisotropy as a function of temperature for FeRh/$BaTiO_3$. An out-of-plane magnetization direction is preferred below the transition temperature.

**6. Resistance and Anomalaous Hall Resistance vs Electric Field for FeRh/$BaTiO_3$**

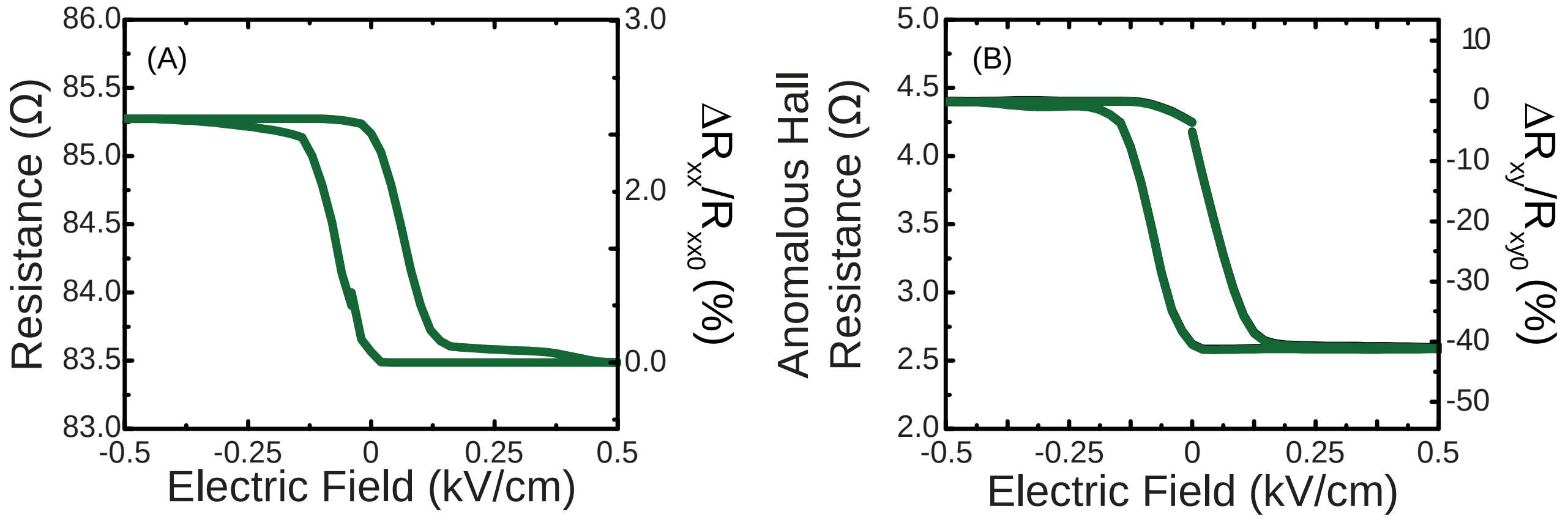

**Fig. S6:** Longitudinal ($R_{xx}$) (A) and transverse ($R_{xy}$, AHR) (B) resistance measurements as a function of electric field for Fe-Rh/$BaTiO_3$ at 60 °C with no external magnetic field. The percent change in the transverse resistance is more than an order of magnitude larger than the percent change in longitudinal resistance indicating that the loop in the transverse resistance is dependent on the out-of-plane magnetization.

### 7. Electric Field Tunability of Longitudinal and Transverse Resistance as a Function of Temperature

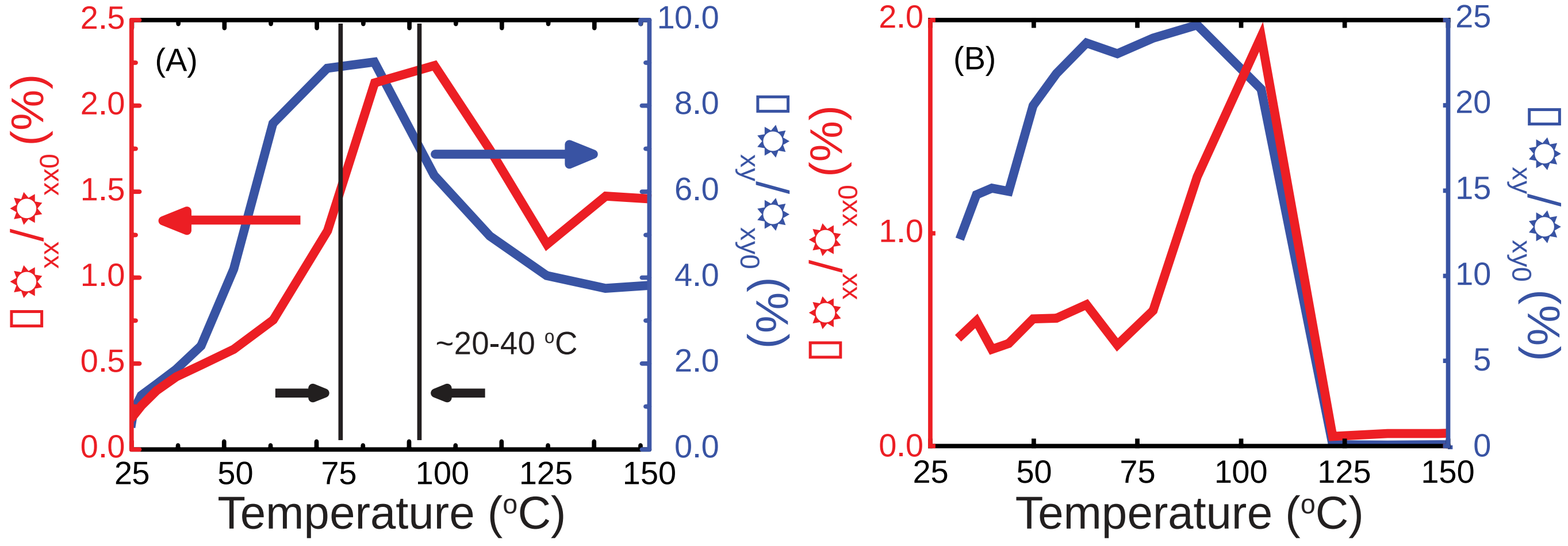

**Fig. S7:** Temperature dependence of electric field control of FeRh longitudinal and transverse resistances for FeRh/PMN-PT (A) and FeRh/$BaTiO_3$ (B). There is a marked shift in the transverse (AHR) modulation to lower temperature as a result of the anisotropy rotation at lower temperatures.

**8. XMCD PEEM of FeRh/PMN-PT**

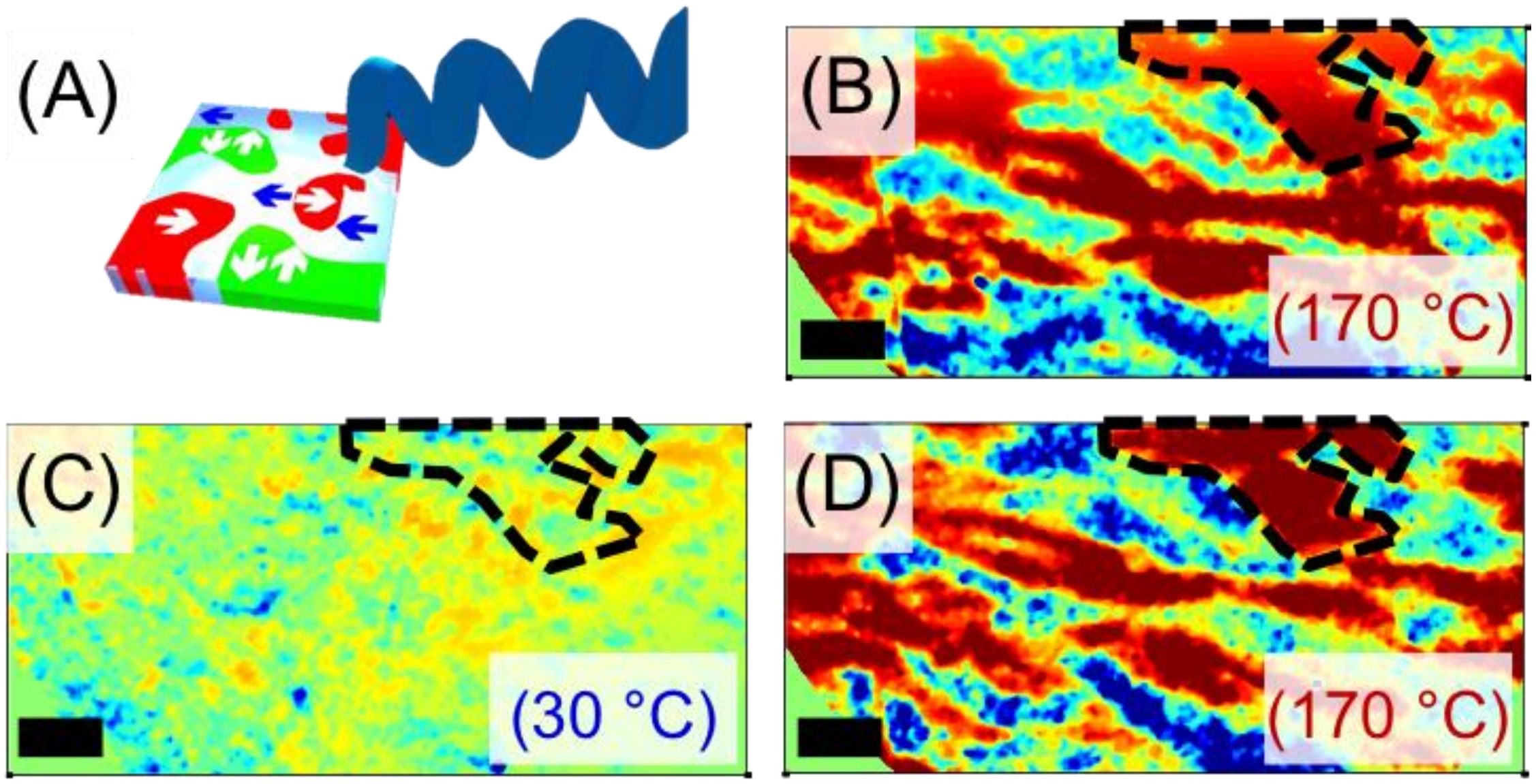

**Fig. S8:** (A) Sketch of the PEEM-XMCD experiment configuration. An x-ray photon (indicated by the blue wave) is directed at a FeRh sample that has a mixture of antiferromagnetic (green) and ferromagnetic (white and red) domains. (B-D) PEEM-XMCD image collected in the very same region at high temperature (170 °C), low temperature (30 °C) and high temperatures (170 °C), respectively, recorded sequentially. The image in (B) has been recorded after saturating the sample at high temperatures, cooling to room temperature in zero field cooling conditions and heating it once. Note that the three images have been collected at magnetic remanence and no magnetic field has been applied during the (B) to (D) measuring sequence. The false color scale corresponds to the projection of the magnetization onto the incident x-ray beam (horizontal from the right). Domains with magnetization parallel and antiparallel to the x-ray incidence have opposite contrast (blue and red colors). Perpendicular domains or domains with zero magnetic moment have green color. The location of the ferromagnetic phase and its magnetization orientation can be reproducibly erased and restored through a heating and cooling cycle as illustrated by the region highlighted by the black dashed line. Scale bar correspond to 1 μm.

## 9. Time and Electric Field Dependence of $FeRh/BaTiO_3$

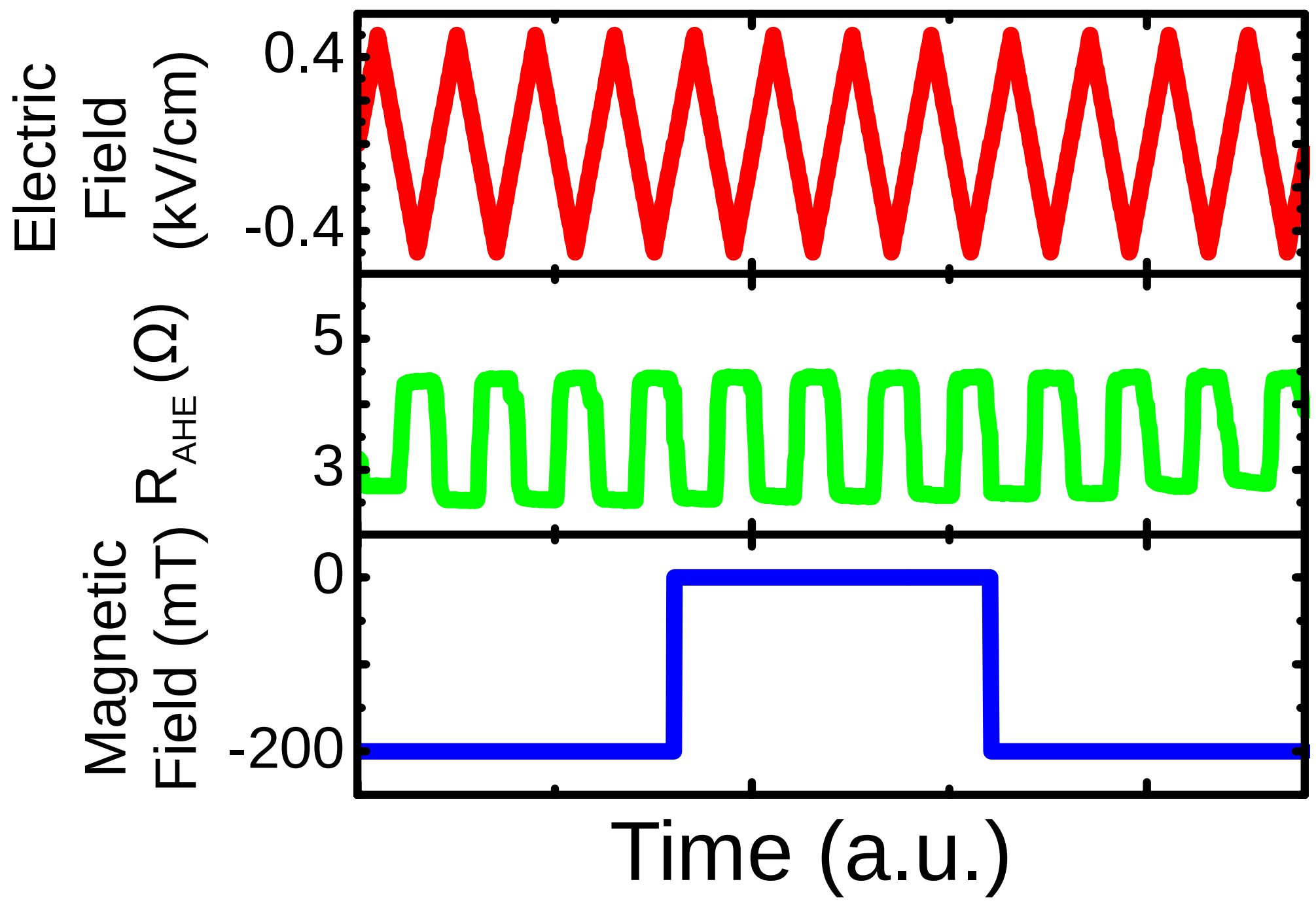

**Fig. S9:** Anomalous Hall resistance as a function of time at negative (-200 mT) and zero magnetic fields through a series of different voltage cycles showing the independence of applied magnetic field on the voltage control of the AHR signal.

## 10. Magnetic Force Microscopy of FeRh/ $BaTiO_3$

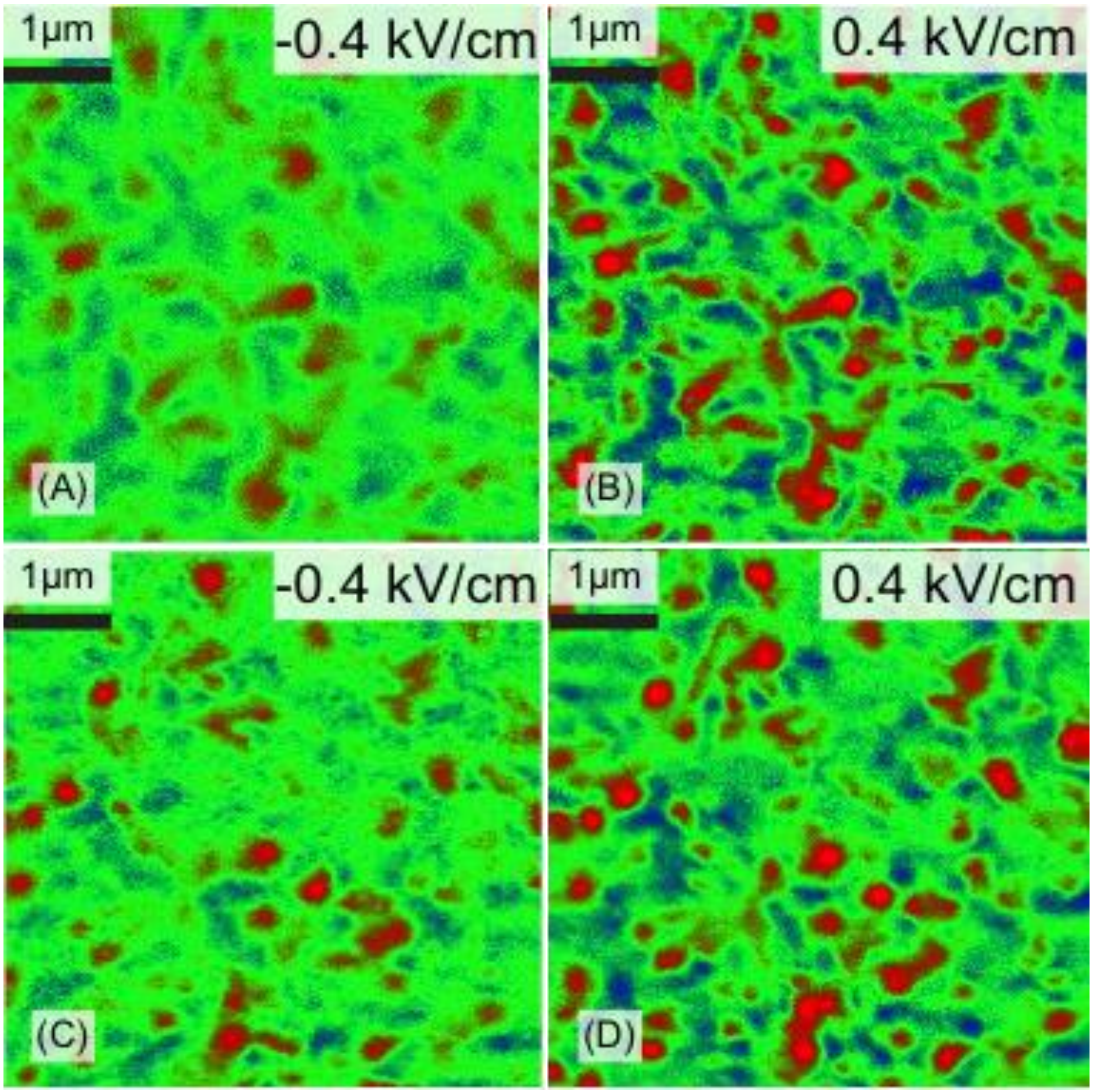

**Fig. S10:** Magnetic force microscopy of the FeRh/$BaTiO_3$ at 60°C after a DC electric field of ±0.4 kV/cm was applied. Images are labelled sequentially with their corresponding electric field. The magnetic response in (A) and (C) are drastically diminished, when compared with the same region in (B) and (D). The local magnetic memory effect is observed here as the local magnetization is reversibly modified with electric field.